\documentstyle[aps,twocolumn]{revtex}
\begin{document}

\title{Orientational pinning of quantum Hall striped phase}
\author{E.~E.~Takhtamirov\thanks{e-mail:gibbon@royal.net} and
V.~A.~Volkov\thanks{e-mail:VoVA@mail.cplire.ru}\\
Institute of Radioengineering and Electronics of RAS,\\
11 Mokhovaya, Moscow 103907, Russia}
\maketitle

\begin{abstract}
In ultra-clean 2D electron systems on (001) GaAs/AlGaAs upon filling high Landau levels, it was
recently observed a new class of collective states, which can be related to the
spontaneous formation of a charge density wave (``striped phase''). We address to the following
unsolved problem: what is the reason for stripe pinning along the crystallographic direction
[110]? It is shown that in a single heterojunction (001) $A_3B_5$ the effective mass of 2D
electrons is anisotropic. This natural anisotropy is due to the reduced ($C_{2v}$) symmetry of the
heterojunction and, even being weak ($0.1\%$), can govern the stripe direction. A magnetic
field parallel to the interface induces ``magnetic'' anisotropy of the effective mass.
The competition of these two types of anisotropy provides quantitative description of the
experiment.
\end{abstract}

%=========================================================================================

\section{Introduction}

It had been assumed \cite{fukuyama} in 1979 that a uniform 2D electron system in strong magnetic
fields corresponding to the filling of the lowest Landau level, $N=0$ (filling factor $\nu < 1$),
can be unstable against the formation of a 1D charge density wave with a period of the order of
magnetic length. This instability is due to the exchange interaction resulting in the effective
attraction between electrons. The analysis was carried in the Hartree-Fock approximation, which
overestimates the exchange interaction and ignores electron-electron correlations.
After the discovery of the fractional quantum Hall effect, it became clear that, for $\nu < 1$,
it is the correlation interaction that leads to the formation of a uniform state of the
Laughlin liquid type. Nevertheless, the role of correlations diminishes with filling a large
number of Landau levels and, in principle, one can expect the indicated instability to appear.
It was predicted in 1996 \cite{fogler} that a 1D charge density wave may appear near the
half-filling of Landau levels, beginning with $N = 2$ \cite{fogler2}. Such a striped phase
with a period of order of the Larmor diameter should be energetically more favorable than
the Laughlin liquid and the Wigner crystal \cite{fogler,fogler2,moessner}.

How should this phase be manifested in the transport measurements if it is really formed
and pinned for some reason? Similar problem was considered, probably, for the first time
in a series of old papers \cite{aizin1,aizin2,aizin3}, where the anisotropic conductivity
was calculated \cite{aizin1} and the high-frequency \cite{aizin2} and heating \cite{aizin3}
effects were studied. In the presence of a periodic 1D potential $U(x)$ induced by the charge
density wave, each Landau level transforms into a narrow 1D band. At the edges of this
band, the density of states has a power divergence, which is cut off upon including weak
scattering. As a result, the density of states $S(E_f)$ at the Fermi level has
the shape of a two-teeth fork: a minimum in the center of the band and two peaks at the
band edges, the lower the density $n_{im}$ of scatterers, the higher the peaks.
Incomplete filling of this band results in the formation of stripes differing in the $\nu$
value [of the type $\nu$/($\nu -1$)/$\nu$/...] and aligned with the $y$-axis.
The transverse ($\sigma_{xx}$) and longitudinal ($\sigma_{yy}$) conductivities obey
different mechanisms and qualitatively differently depend on $\nu$: the $\sigma_{yy}$
conductivity is high, has the band character, and is inversely proportional to $n_{im}$ and
$S^2(E_f)$, while $\sigma_{xx}$ is low, has the hopping character, and is proportional to
$n_{im}$ and $S^2(E_f)$, with the product $\sigma_{xx}\sigma_{yy}$ being independent of
scattering. These results were in fact confirmed and generalized in \cite{mcdonald,galperin}.
What did the experiment actually reveal?

In 1999, the conductivity of an electron system with ultrahigh mobility in the (001)
GaAs/AlGaAs structures was studied at very low temperatures near half-integer $\nu \geq 9/2$,
and a new state was revealed \cite{lilly,du,pan,lilly2} which was assumed to be just the one
associated with the formation of the striped phase predicted in \cite{fogler}.
This assumption was primarily based on the observation of a giant resistance anisotropy
in this system. The ratio of resistances along the crystallographic directions $[1\bar 10]$ and
$[110]$ reaches the values of $R_{xx}/R_{yy}\sim5$-$3500$, depending on the sample geometry,
where $[110]$ is the ``easy'' conductivity direction. Moreover, the behavior of all conductivity
tensor components qualitatively agrees with the theory \cite{aizin1,mcdonald,galperin}.
The predicted behavior of the $\sigma_{xx} \sigma_{yy}$ product near the half filling of the
upper Landau level numerically agrees with the experiment \cite{new-eisen}. It was also shown
in \cite{pan,lilly2} that the magnetic field $B_{\|}\sim 1$\,T parallel to the interface can
change the direction of easy conductivity. The authors of \cite{pan} concluded that, at high
enough $B_{\|}$, the direction of easy conductivity is perpendicular to the $B_{\|}$
direction. Similar result was obtained in \cite{lilly2} for ${\bf B}_{\|} \| [110]$ near all
half-integer $\nu \geq 9/2$ and for ${\bf B}_{\|} \| [1\bar 10]$ near $\nu = 11/2$ and
$\nu = 15/2$. The theoretical analysis \cite{jungwirth,stanescu} of the influence of
${\bf B}_{\|}$ carried out in the Hartree-Fock approximation in the model of parabolic
quantum well partially explained the results. All this is strong evidence for the formation
of a striped phase. However, a mechanism responsible for the orientation of the charge
density stripes along a certain preferred direction in a macroscopic sample
(orientational pinning) at $B_{\|}=0$ remains to be clarified. This is
one of the fundamental unsolved problems.

In this work, the Kroemer's assumption \cite{kroemer} that the reduced ($C_{2v}$)
symmetry of the potential in a heterostructure based on semiconductors without inversion
center can cause an appearance of a preferred direction for the conductivity is justified.
Symmetry reduction means that the cubic axis normal to the interface is transformed from
the fourfold mirror-rotational axis ($S_4$) to the twofold axis ($C_2$). We demonstrate below
that, owing to the asymmetry of the potential of an atomically sharp heterojunction, the
effective mass (EM) of 2D electrons is anisotropic (natural anisotropy). At the same time,
the presence of ${\bf B}_{\|}$ also gives rise to the EM anisotropy (magnetic anisotropy)
\cite{ando}. Therefore, the results of many-particle numerical calculations
of the ${\bf B}_{\|}$ effect on the orientation of striped phase \cite{jungwirth,stanescu}
is natural to treat (to the lowest order in $B^2_{\|}$) as a manifestation of the magnetic
anisotropy of EM. Thus, the many-particle problem of orientational pinning of the striped
phase reduces to a one-particle problem of determining the EM anisotropy. We derive below
analytical expressions for both types of EM anisotropy (natural and magnetic) and
demonstrate that they can compete with each other. At a certain magnitude and direction of
${\bf B}_{\|}$, these two types of anisotropy exactly cancel, leading to the disappearance of
resistance anisotropy, in agreement with the experiment.

\section{Natural anisotropy}

Before proceeding to the many-particle problem, it is necessary to obtain one-particle
Hamiltonian for the conduction band in a (001) $A_3B_5$ heterostructure. As was shown in
\cite{we}, the correctly constructed multiband set of equations for the envelope functions
retains information on the heterostructure symmetry ($C_{2v}$), which is lower than the
symmetry $T_d$ of the constituent materials. This symmetry reduction is described, in
particular, by certain short-range potentials localized at the heterointerface. Mixing of
heavy and light holes at the center of the 2D Brillouin zone is one of the consequences of
symmetry reduction \cite{we,ivchenko}. This mixing explains giant optical anisotropy
(with the same principal axes $[110]$ and $[1\bar 10]$) that was discovered in \cite{krebs}
for the quantum wells based on semiconductors with different cations and anions. Evidently,
low symmetry should also manifest itself in the equation for the envelope functions in
the conduction band. Nevertheless, a single-band equation obtained in \cite{we} carries no
information on the $C_{2v}$ symmetry because the corresponding small contributions were
neglected. We must now take them into account. Since the terms with the symmetry higher than
$C_{2v}$ are of no interest here, the effective Hamiltonian can include only the operators
of kinetic and potential energies used in the standard EM approximation, as well as the
anisotropic contribution of $C_{2v}$ symmetry, which will be obtained below.

A single-band Hamiltonian of the $C_{2v}$ symmetry can be obtained by the method of invariants.
Leading aside spin-orbit interaction, one can conclude that the $C_{2v}$ symmetry should
manifest itself in the kinetic energy operator. Let us direct the 2D quasi-momentum
components along the cubic axes: $p_x \| [100]$ and $p_y \| [010]$ ($z$ axis is along $[001]$).
Then the part of kinetic energy operator quadratic in the generalized 2D momentum $(P_x\,,P_y)$
should be
%================================================================================
\begin{eqnarray}
T = \frac {P^2_x + P^2_y}{2m^*} +\frac 12 {\cal A} \left(P_x P_y + P_y P_x\right).
\label{kin}
\end{eqnarray}
%================================================================================
Here, $m^*$ is the EM of conduction band, and the quantity $\cal A$ (which may depend on $z$)
accounts for the natural anisotropy of EM in the plane of 2D electron gas. Let us obtain
explicit expression for ${\cal A}$ by using a multiband matrix Hamiltonian \cite{we}
acting on the column of envelope functions. In the $\bf k$ representation, it takes the form
%================================================================================
\begin{eqnarray}
H^{\left( eff\right)}_{nn^{\prime}}= H^{\left( kp\right)}_{nn^{\prime}}
+\frac 1{2\pi}D_{0nn^{\prime}},
\label{ham-kp-d}
\end{eqnarray}
%================================================================================
where $n$ and $n^{\prime}$ are the band indices. The first term in Eq.~(\ref{ham-kp-d})
includes the contributions from the smooth potentials and $\bf kp$ interaction and has the
standard form. The second term in Eq.~(\ref{ham-kp-d}) is a contribution from the atomically
sharp heterointerface potential taken to a first order in the small parameter $\bar ka$,
where $1/\bar k$ is the characteristic length of changing the envelope functions and $a$
is the lattice constant. One can pass to the single-band variant of envelope-function
method by applying the perturbation theory, with the $\bf kp$ and ${\bf D}_{0}$ operators as
perturbation. The second order in the $\bf kp$ interaction gives the first (standard)
term in Eq.~(\ref{kin}). The third order (second order in $\bf kp$ and first in
${\bf D}_{0}$) provides the second term of Eq.~(\ref{kin}), with
${\cal A}(z) =\alpha \delta (z)$ and
%================================================================================
\begin{eqnarray}
\alpha &=& {\sum_{n,n^{\prime}}}^{\prime }
\frac{
2 \left\langle c\mid p_x \mid n\right\rangle D_{0nn^{\prime}}
\left\langle n^{\prime} \mid p_y \mid c\right\rangle
}
{
m_0^2\left( \epsilon _{c}-\epsilon_{n}\right)
\left( \epsilon _{c}-\epsilon _{n^{\prime}}\right)
}\nonumber\\
& & + {\sum_{n,n^{\prime}}}^{\prime }
\frac{
4 D_{0cn} \left\langle n \mid p_x \mid n^{\prime} \right\rangle
\left\langle n^{\prime} \mid p_y \mid c \right\rangle
}
{
m_0^2\left( \epsilon _{c}-\epsilon_{n}\right)
\left( \epsilon _{c}-\epsilon _{n^{\prime}}\right)
}.
\label{alfa}
\end{eqnarray}
%=============================================================================
Here $\delta (z)$ is the Dirac $\delta$-function, $z = 0$ defines the heterointerface
position, $\left\langle n \mid p_i \mid n^{\prime} \right\rangle$ is the $i$th component
of the interband momentum matrix element, $c$ is the index of conduction band, $m_0$ is the
mass of free electron, and $\epsilon _{n}$ is the energy of the $n$th band edge in one of
the structure materials. In the simplest model, the key parameters of the theory
($D_{0nn^{\prime}}$) have the form
%================================================================================
\begin{eqnarray}
D_{0nn^{\prime}}&=&\sum_{j=\pm 1, \pm 2, ...}
\frac {\left\langle n\mid \delta U\sin(4\pi jz/a) \mid n^{\prime}\right\rangle } {4\pi j/a}
\nonumber\\
& & \times \int\limits_{-\infty}^{+\infty} \frac {dG(z)}{dz}
\cos \left( \frac {4\pi}{a}jz\right) dz .
\end{eqnarray}
%================================================================================
The functions $G(z)$ and $\delta U({\bf r})$ are so defined that the crystal potential
of the heterostructure has the form $U({\bf r}) = U_1({\bf r}) + G(z) \delta U({\bf r})$,
where $U_1$ and $U_2 = U_1 + \delta U$ are the crystal potentials of the structure materials.
Note that the $D_{0XY}$ parameter accounts for the mixing of the heavy and light holes at the
center of the 2D Brillouin zone, with $X$ and $Y$ being the indices of the Bloch functions
corresponding to the edge of the $\Gamma_{15}$ valence band and transforming as $x$ and $y$
under symmetry operations of the $T_d$ group\cite{we,ivchenko}.

\section{Inclusion of the magnetic anisotropy}

Reducing the tensor of reciprocal effective mass to the principal axes, so that
$x \| [1 \bar 1 0]$ and $y \| [110]$ in the new coordinates, and introducing magnetic field
$\bf B$ in the vector-potential gauge ${\bf A} = (B_y z, -B_x z + B_z x, 0)$, one obtains for
the orbital part of the 3D Hamiltonian of conduction band
%===============================================================================
\begin{eqnarray}
H_{3D} &=& V(z) + \frac {p^2_z}{2m^*} +
\frac 12 \left( \frac 1 {m^*} - \alpha \delta (z) \right)
\left( p_x + \frac e c B_y z \right)^2\nonumber\\
&&+\frac 12 \left( \frac 1 {m^*} + \alpha \delta (z) \right)
\left( p_y - \frac e c B_x z + \frac e c B_z x \right)^2.\label{h}
\end{eqnarray}
%===============================================================================
Here, $V(z)$ is the effective potential of the conduction band edge, $e$ is the elementary
charge, and $c$ is the light speed (we hope, it will be no confusion with the conduction band
index). For the finite thickness of the 2D layer, the magnetic field component parallel to the
heterointerface can be treated perturbatively \cite{ando}. To second order in $B_{\|}$,
this procedure brings about a diamagnetic shift of the dimensional-quantization subbands and
an increase (for the lowest subband) in EM in the direction perpendicular to ${\bf B}_{\|}$.
The natural EM anisotropy also can be treated perturbatively. For simplicity, we assume that
${\bf B}_{\|}$ is parallel to either $[1 \bar 1 0]$ or $[110]$, so that $B_xB_y = 0$.
Collecting all terms second-order in $B_{\|}$ and first-order in $\alpha$, one obtains the
following expression for the 2D Hamiltonian of the lowest subband:
%===============================================================================
\[
H ^1_{2D} = E_1 + \frac {e^2}{2m^*c^2} \left( B^2_x + B^2_y \right)
\left(\left\langle z^2 \right\rangle _{11} - \left\langle z \right\rangle _{11}^2\right)
\]
\begin{equation}
+\frac 1{2m^*} \left[ 1 - \frac {\Delta _{nat}}2 - \frac {B_y^2}{B_{\|}^2} \Delta _B \right]\!\!
\left( p_x + \frac e c B_y \left\langle z \right\rangle _{11} \right)^2 \label{h0}\\
\end{equation}
\[
+\frac 1{2m^*} \left[ 1 + \frac {\Delta _{nat}}2 - \frac {B_x^2}{B_{\|}^2} \Delta _B \right]
\!\!
\left( p_y +\frac e c B_z x -\frac e c B_x \left\langle z \right\rangle _{11} \right)^2 .
\]
%===============================================================================
The parameters of the natural EM anisotropy and the EM anisotropy induced by the magnetic
field are
%===========================================================================
\begin{eqnarray}
\Delta_{nat} = 2 m^*\alpha\left\langle \delta (z) \right\rangle _{11},\ 
\Delta_B = \frac {2e^2B^2_{\|}}{m^*c^2}{\sum_m}^{\prime }
\frac {\mid \left\langle z \right\rangle _{1m}\mid ^2}{E_m - E_1},\label{delty}
\end{eqnarray}
%===========================================================================
where $E_m$ is the energy of the bottom of the $m$th subband at $B=0$. The expression for
$\Delta_B$ in Eq.~(\ref{delty}) is valid to the terms second-order in the parameter
$\hbar\omega_c/(E_2 - E_1)$, where $\omega_c=eB_z/m^*c$. For the field $B_z=2.5$\,T
(in the experiment \cite{lilly2} this field corresponds to the filling
factor $\nu=9/2$), one can neglect this correction in the estimation of $\Delta_B$,
because $\hbar\omega_c\approx 4$\,meV, while the gap $E_2 - E_1$ should exceed the
Fermi energy $E_f$ measured from the lowest subband; one has $E_f \approx 10$\,meV for the
2D electron concentration $N_s=2.7\times 10^{11}$\,cm$^{-2}$.

\section{Estimates}

Based on the experimental data \cite{lilly2}, we estimate $\Delta_{nat}$ and $\Delta_B$ for
$B_{\|} = 0.5$\,T (if ${\bf B}_{\|} \| [110]$, this magnetic field converts the resistance
from anisotropic to isotropic; at larger $B_{\|}$ the direction of ``easy'' conductivity
rotates by $90^\circ$). Since the information on the samples is incomplete, we carried out
a series of self-consistent calculations by varying the concentration $N_a$ of residual
acceptors in GaAs. At $N_a=10^{14}$\,cm$^{-3}$ and $N_s$ taken from \cite{lilly2},
the results are
%===========================================================================
\begin{eqnarray}
{\sum_m}^{\prime }\frac {\mid \left\langle z \right\rangle _{1m} \mid ^2}{E_m - E_1}
&\approx& 1\times 10^{-11}\,cm^2/eV,\label{otsenka1}\\
\left\langle \delta (z) \right\rangle _{11} &\approx& 1\times 10^{5}\,cm^{-1}.
\label{otsenka2}
\end{eqnarray}
%============================================================================
For other $N_a$ values (from $10^{13}$ to $10^{15}$\,cm$^{-3}$), the results differ from
Eq.~(\ref{otsenka1}) and ~(\ref{otsenka2}) by a factor less than two. Equations (\ref{delty})
and (\ref{otsenka1}) yield the following value for the EM anisotropy induced by magnetic field
$B_{\|}=0.5$\,T:
%===========================================================================
\begin{eqnarray}
\Delta_B = 1.3 \times 10^{-3} = 0.13\%.
\end{eqnarray}
%============================================================================
The parameter $\alpha$ can be determined from the equality $\Delta_{nat}=\Delta_B$ to give
%===========================================================================
\begin{eqnarray}
\alpha = \frac {0.65\times 10^{-8}\,cm}{m^*} = 1.1 \times 10^{20}\,cm/g.
\label{alfa-value}
\end{eqnarray}
%============================================================================
The two-band approximation with energy gap $E_g$ yields the following estimate for
Eq.~(\ref{alfa}):
%============================================================================
\begin{eqnarray}
\alpha \sim
\frac{2 \left\langle c\mid p_x \mid X\right\rangle D_{0XY}
\left\langle Y \mid p_y \mid c\right\rangle}{m_0^2 E_g^2}
= \frac {D_{0XY}}{m^*E_g}. \label{2band}
\end{eqnarray}
%============================================================================
Thus, it follows from the experimental data \cite{lilly2} and Eqs.~(\ref{alfa-value})
and (\ref{2band}) that $D_{0XY}\sim 0.4\times 10^{-8}$\,eV\,cm. Let us compare this value
with the literature data.

It was found in \cite{ivchenko} that different estimates carried out for the
GaAs/AlAs heterostructures either on the basis of pseudopotential and tight-binding
calculations or from the comparison with the experiment on the anisotropic exchange
splitting of exciton levels in the II-type superlattices GaAs/AlAs lead to a sizable
scatter of the $D_{0XY}$ parameter. The value obtained in \cite{ivchenko} lies in the range
$(0.35,\;0.99)\times 10^{-8}$\,eV\,cm. Using linear interpolation, one obtains the upper bound
$D_{0XY}=0.3\times 10^{-8}$\,eV\,cm for the $\rm GaAs/Al_{0.3}Ga_{0.7}As$ heterostructure.
This value is in a fair agreement with the value obtained above.

\section{Discussion}

We can now conclude that the natural anisotropy of EM is likely the mechanism that pins the
stripe directions at $B_{\|}=0$ (see also the end of section 1). It follows from this
conclusion that the parameter $\alpha$ entering Eq.~(\ref{kin}) is negative, $\alpha < 0$.
The competition between the natural anisotropy $\Delta_{nat}$ and anisotropy $\Delta_B$
induced by the magnetic field ${\bf B}_{\|} = (0,\, B_y)$ makes the 2D electron spectrum
at $B_{\|}=0.5$\,T isotropic. As a result, the stripe directions are randomized and the
resistance becomes isotropic. On further increase in $B_{\|}$, the magnetic anisotropy
prevails and the stripes rotate at $90^\circ$. The role of EM anisotropy in the formation
of many-electron anisotropic states can be understood as follows. A 2D electron system with
anisotropic EM and isotropic Coulomb interaction is, obviously, equivalent to a 2D electron
system with isotropic (cyclotron) mass and anisotropic Coulomb interaction. One can expect
that this effective anisotropic interaction is precisely the one which pins the orientation
of the striped phase to ensure its observation in the magnetotransport.

For holes, the heterointerface contribution of symmetry $C_{2v}$ (and, hence, responsible
for the pinning of the striped phase) is greater than for the conduction band, because it
appears in the first-order perturbation treatment \cite{we}, whereas the anisotropic EM
in Eq.~(\ref{kin}) was obtained in the third order. For this reason, one would expect that
the hole striped phase is more stable and can form upon filling the lower Landau levels
(cf.~\cite{shayegan}).

\section*{Acknowledgement}

We are grateful to B. I. ShklovskiÏ for stimulating discussion and useful remark.
%======================================================================


\begin{thebibliography}{99}

\bibitem{fukuyama} H.~Fukuyama, P.~M.~Platzman, and P.~W.~Anderson, Phys.\ Rev. {\bf B19}
(1979) 5211.

\bibitem{fogler} M.~M.~Fogler, A.~A.~Koulakov, and B.~I.~Shklovskii, Phys.\ Rev. {\bf B54}
(1996) 1853.

\bibitem{fogler2} M.~M.~Fogler and A.~A.~Koulakov, Phys.\ Rev. {\bf B55} (1997) 9326.

\bibitem{moessner} R.~Moessner and J.~T.~Chalker, Phys.\ Rev. {\bf B54} (1996) 5006.

\bibitem{aizin1} G.~R.~Aizin and V.~A.~Volkov, Zh.\ Eksp.\ Teor.\ Fiz. {\bf 87} (1984) 1469
[Sov.\ Phys.\ JETP {\bf 60} (1984) 844]; Zh.\ Eksp.\ Teor.\ Fiz. {\bf 92} (1987) 329
[Sov.\ Phys.\ JETP {\bf 65} (1987) 188].

\bibitem{aizin2} G.~R.~Aizin and V.~A.~Volkov, Fiz.\ Tekh.\ Poluprovodn. (Leningrad)
{\bf 19} (1985) 1780 [Sov.\ Phys.\ Semicond.\ {\bf 19} (1985) 1094].

\bibitem{aizin3} G.~R.~Aizin and V.~A.~Volkov, Fiz.\ Tverd.\ Tela (Le\-nin\-grad) {\bf 27} (1985) 475
[Sov.\ Phys.\ Solid State {\bf 27} (1985) 290].

\bibitem{mcdonald} A.~H.~MacDonald and M.~P.~A.~Fisher, Phys.\ Rev. {\bf B61} (2000) 5724.

\bibitem{galperin} F.~von~Oppen, B.~I.~Halperin, and A.~Stern, Phys.\ Rev.\ Lett. {\bf 84} (2000)
2937.

\bibitem{lilly} M.~P.~Lilly, K.~B.~Cooper, J.~P.~Eisenstein, L.~N.~Pfeif\-fer, and K.~W.~West,
Phys.\ Rev.\ Lett. {\bf 82} (1999) 394.

\bibitem{du} R.~R.~Du, D.~C.~Tsui, H.~L.~Stormer, L.~N.~Pfeiffer, K.~W.~Baldwin, and K.~W.~West,
Solid State Commun. {\bf 109} (1999) 389.

\bibitem{pan} W.~Pan, R.~R.~Du, H.~L.~Stormer, D.~C.~Tsui, L.~N.~Pfeiffer, K.~W.~Baldwin, and
K.~W.~West, Phys.\ Rev.\ Lett. {\bf 83} (1999) 820.

\bibitem{lilly2} M.~P.~Lilly, K.~B.~Cooper, J.~P.~Eisenstein, L.~N.~Pfeiffer, and K.~W.~West,
Phys.\ Rev.\ Lett. {\bf 83} (1999) 824.

\bibitem{new-eisen} J.~P.~Eisenstein, M.~P.~Lilly, K.~B.~Cooper, L.~N.~Pfeiffer, and K.~W.~West,
cond-mat/0003405.

\bibitem{jungwirth} T.~Jungwirth, A.~H.~MacDonald, L.~Smr\v cka, and S.~M.~Gir\-vin,
Phys.\ Rev. {\bf B60} (1999) 15574.

\bibitem{stanescu} T.~D.~Stanescu, I.~Martin, and P.~Phillips, Phys.\ Rev.\ Lett. {\bf 84} (2000)
1288.

\bibitem{kroemer} H.~Kroemer, cond-mat/9901016.

\bibitem{ando} T.~Ando, A.~Fowler, and F.~Stern, Rev.\ Mod.\ Phys. {\bf 54} (1982) 437.

\bibitem{we} E.~E.~Takhtamirov and V.~A.~Volkov, Zh.\ Eksp.\ Teor.\ Fiz.\ {\bf 116} (1999) 1843
[JETP {\bf 89} (1999) 1000].

\bibitem{ivchenko} E.~L.~Ivchenko, A.~Yu.~Kaminski, and U.~R\"ossler, Phys.\ Rev. {\bf B54} (1996)
5852.

\bibitem{krebs} O.~Krebs, W.~Seidel, J.~P.~Andr\'e, D.~Bertho, C.~Jouanin, and P.~Voisin,
Semicond.\ Sci.\ Technol. {\bf 12} (1997) 938.

\bibitem{shayegan} M.~Shayegan, H.~C.~Manoharan, S.~J.~Papadakis, and E.~P.~De~Poortere,
Physica E, {\bf 6} (2000) 40.

\end{thebibliography}
\end{document}